\documentclass{article} 
\usepackage{iclr2025_conference,times}

\usepackage[utf8]{inputenc} 
\usepackage[T1]{fontenc}    
\usepackage{hyperref}       
\usepackage{url}            
\usepackage{booktabs}       
\usepackage{amsfonts}       
\usepackage{nicefrac}       
\usepackage{microtype}      
\usepackage{graphicx}
\usepackage{amsmath}

\usepackage{xcolor}
\usepackage{adjustbox}

\usepackage{algorithm}
\usepackage{algpseudocode}
\usepackage{booktabs}
\usepackage{makecell}

\iclrfinalcopy

\usepackage{todonotes}

\title{Non-Equilibrium Dynamics of Hybrid \\ Continuous-Discrete Ground-State Sampling}

\author{
  Timothée Leleu \\
  NTT Research, Stanford University \\
  \texttt{timothee.leleu@ntt-research.com}\\
  \texttt{tleleu@stanford.edu}
  \And
  Samuel Reifenstein \\
  NTT Research \\
  \texttt{samuel.reifenstein@ntt-research.com}
}

\usepackage{titlesec}
\newcommand{\supplementprefix}{S}

\newcommand{\beginsupplement}{
    \renewcommand\thesection{\supplementprefix\arabic{section}}
    \renewcommand{\theequation}{\supplementprefix\arabic{equation}}
    \renewcommand\thefigure{\supplementprefix\arabic{figure}}
    \setcounter{section}{0}
    \setcounter{equation}{0}
    \setcounter{figure}{0}
}

\begin{document}

\maketitle

\begin{abstract}
We propose a general framework for a hybrid continuous-discrete algorithm that integrates continuous-time deterministic dynamics with Metropolis-Hastings steps to combine search dynamics with and without detailed balance. Our purpose is to study the non-equilibrium dynamics that leads to the ground state of rugged energy landscapes in this general setting. Our results show that MH-driven dynamics reach ``easy'' ground states faster, indicating a stronger bias in the non-equilibrium dynamics of the algorithm with reversible transition probabilities. To validate this, we construct a set of Ising problem instances with a controllable bias in the energy landscape that makes one degenerate solution more accessible than another. The constructed hybrid algorithm demonstrates significant improvements in convergence and ground-state sampling accuracy, achieving a 100x speedup on GPUs compared to simulated annealing, making it well-suited for large-scale applications.
\end{abstract}

\section{Introduction}

The fact that non-equilibrium relaxation dynamics reach different degenerate ground states at different rates, contrary to what is expected at equilibrium, is a fundamental topic in optimization and statistical physics~\citep{Bernaschi2020,Lucas2014,Mohseni2022}. In machine learning, recent works suggest that sampling from more easily accessible degenerate ground states favors finding models that generalize well during the training of overparameterized neural networks~\citep{Soudry2018,baity2018comparing,feng2021inverse,baldassi2022learning,baldassi2023typical}. Here, we study this issue from the general perspective of non-convex optimization, focusing on the Ising Hamiltonian. In particular, we study the issue of sampling from degenerate ground states in non-convex energy landscapes, specifically the distinction between Boltzmann and chaotic sampling characterized by detailed balance (reversible) and asymmetric transition probabilities (non-reversible), respectively.

The Metropolis-Hastings (MH) algorithm is one of the most widely used techniques for sampling from a reversible Markov Chain Monte Carlo (MCMC)~\citep{Zhang2020,Chib1995,Chen2014}. This approach proposes new states and accepts them with a probability designed to ensure that the stationary distribution of the chain is the Boltzmann distribution. Despite its simplicity and versatility, MH can struggle with getting trapped in local energy minima, especially when dealing with complex energy landscapes. Examples of these energy landscape often arise in systems exhibiting broken ergodicity~\citep{Bernaschi2020}, such as spin glasses. In these cases, other MCMC algorithms that maintain detailed balance, like simulated annealing~\citep{kirkpatrick1983optimization} and more advanced methods~\citep{hukushima1996exchange,surjanovic2022parallel,hukushima2003population,houdayer2001cluster,landau2004new}, are often favored for exploring rugged energy landscapes and finding optimal solutions. Interestingly, several algorithms do not satisfy detailed balance yet remain competitive~\citep{Leleu2019,Leleu2021,goto2021high,buesing2011neural}, particularly those based on chaotic sampling~\citep{Leleu2019,Leleu2021}. Although these approaches do not sample in general from the exact Boltzmann distribution at equilibrium, it can be shown in special cases that non-reversible Markov chains mix faster than reversible ones~\citep{kapfer2017irreversible}. However, their ability to reach degenerate ground states of rugged energy landscape during non-equilibrium dynamics is still not well understood.

In this work, we introduce a generalized algorithm that combines MCMC techniques through a MH step with chaotic search dynamics via the chaotic amplitude control (CAC) algorithm. The hyperparameters of this hybrid approach allow for a smooth transition between the two strategies. Additionally, we generate new planted instances with degenerate ground states, inspired by the Wishart planted models. A tunable parameter controls the bias towards different degenerate ground states, enabling us to investigate the non-equilibrium behavior of the hybrid MH/chaotic algorithm in locating solutions that vary in difficulty. We demonstrate that the algorithm's performance can be optimized based on the instance bias, making it competitive with state-of-the-art methods.

\section{Related Work}

\subsection{Discrete state MCMC optimizers}

MCMC methods for optimizing energy functions over discrete spaces—such as simulated annealing ~\citep{kirkpatrick1983optimization} and parallel tempering ~\citep{hukushima1996exchange}—are standard approaches in the field of optimization. Building upon these, recent schemes such as population annealing ~\citep{hukushima2003population}, variational parallel tempering ~\citep{surjanovic2022parallel} and specialized techniques, including cluster updates ~\citep{houdayer2001cluster}, Wang-Landau sampling ~\citep{landau2004new}, and Replica Exchange Wang-Landau Sampling~\citep{vogel2013generic}, have been developed to enhance convergence and sampling efficiency. However, these methods often struggle in higher-dimensional discrete spaces due to the complexity of the energy landscape and the tendency to become trapped in local energy minima caused by broken ergodicity, which is a behavior observed in many systems with frustrated interactions.~\citep{Bernaschi2020}.

\subsection{Hybrid continuous-discrete samplers}

For continuous space sampling, a plethora of methods have been developed based on the idea of combining continuous-time dynamics (typically via ODEs or SDEs) with discrete MCMC corrections. Some examples of this are Metropolis Adjusted Langevin Dynamics (MALA) ~\citep{Roberts1998}, Hamiltonian Monte Carlo (HMC) ~\citep{neal2011mcmc,Xifara2014}, as well as derivative methods such as the No-U-Turn Sampler (NUTS) ~\citep{Hoffman2014}, Generalized HMC ~\citep{Chen2014}, HMC with the Gaussian integral trick ~\citep{zhang2012continuous}, and amortized Metropolis adjustments ~\citep{Zhang2020}. These methods are primarily designed for continuous spaces and do not directly apply to discrete-state sampling. However, research has been conducted to extend continuous sampling techniques to discrete spaces, including methods like discontinuous HMC ~\citep{nishimura2020discontinuous}, Gibbs with Gradient (GWG) ~\citep{Grathwohl2021}, and the Hamming Ball Sampler ~\citep{titsias2017hamming}. Nonetheless, these approaches are not suitable for ground-state sampling.

\subsection{Hybrid continuous-discrete optimizers}

Recently, there has been renewed interest in using continuous-space ODEs to find ground states in discrete spaces. This effort has been driven primarily by the idea of leveraging a diverse array of physical hardware, including CMOS, memristors, spintronics, superconducting circuits, and photonics (see review ~\citep{Mohseni2022}), all designed to solve optimization problems more efficiently than current high-performance digital computers.

In particular, dynamical systems such as simulated bifurcation machines ~\citep{goto2021high}, analog iterative machines (AIM) ~\citep{Kalinin2023}, and chaotic amplitude control (CAC) ~\citep{Leleu2019,Leleu2021,reifenstein2023coherent} offer an alternative approach. Rather than relying on gradient descent of an energy function in systems with symmetric interactions, such as in the seminal work by Hopfield ~\citep{Hopfield1985}, it has been shown that Hamiltonian dynamics ~\citep{goto2021high,neal2011mcmc} or chaotic dynamics driven by asymmetric interactions ~\citep{Leleu2019,Leleu2021} can outperform standard heuristics like simulated annealing ~\citep{kirkpatrick1983optimization}, parallel tempering ~\citep{hukushima2003population}, and breakout local search ~\citep{BenlicBLS} on specific benchmark datasets ~\citep{Leleu2021}, challenging the prevailing belief that MCMC heuristics maintain an advantage in discrete optimization over systems whose dynamics are relaxed into the continuous domain.

Hybrid approaches that combine continuous-space ODEs with discrete MCMC probabilistic transitions remain underexplored in the context of ground-state search, as opposed to their more common application in Boltzmann sampling.


\section{Method}

\subsection{Overview}

\begin{figure}[h]
  \centering
\includegraphics[width=0.90\columnwidth]{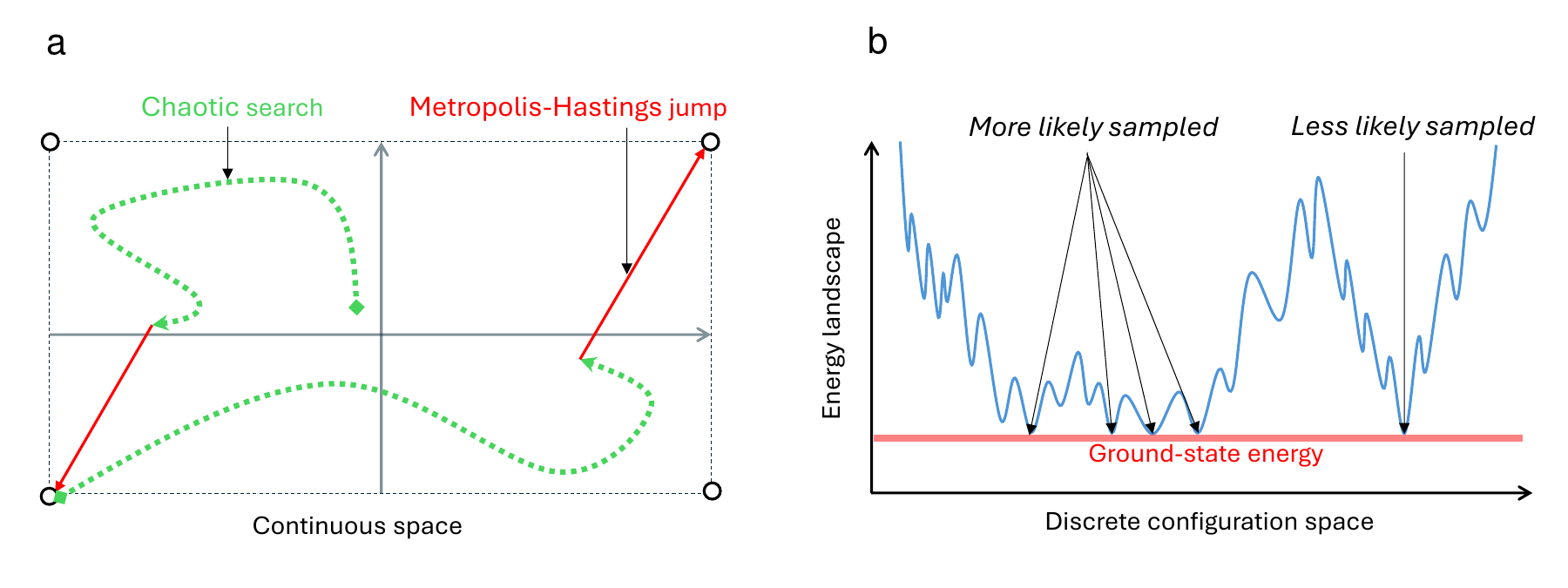}
  \caption{\label{fig:1} a) Proposed hybrid approach generates deterministic chaotic trajectories that rapidly explores the energy landscape and probabilistic jumps accepted via the Metropolis-Hastings criterion. (b) Constructed planted instances with degenerate ground states with unequal sampling probabilities by heuristics. Some ground states are easier and others harder to be find by a local search heuristic.}
\end{figure}

In this section, we introduce a novel MCMC method that combines chaotic search with the MH algorithm~\citep{hastings1970} (see Fig.~\ref{fig:1} a). The algorithm is built on three main concepts: (1) relaxation from discrete to continuous space (or continuous embedding), (2) addition of auxiliary variables to escape from local minima (chaotic amplitude control~\citep{Leleu2019,Leleu2021} and momentum~\citep{Kalinin2023}), (3) probabilistic jumps in the discrete space based on the Metropolis-Hastings criterion.

In the next section, we construct a novel set of Ising problem instances inspired from the Wishart planted ensemble~\citep{hamze2020wishart} in order benchmark this algorithm's ability to sample from optimal solutions (or ground-state energies). The constructed instances exhibit degenerate ground states that are sampled with unequal probabilities by local search algorithms (see Fig.~\ref{fig:1} b).

\subsection{Deterministic continuous time model}

We define a continuous time dynamical system inspired by the CAC algorithm~\citep{Leleu2019,Leleu2021} and generalize its formulation to draw a correspondence between its parameters and the inverse temperature $\beta$ of the Boltzmann distribution $P(\boldsymbol{\sigma}) \propto e^{\beta V(\boldsymbol{\sigma})}$. The aim is to construct a Markov chain that is capable of sampling from the distribution $P(\boldsymbol{\sigma})$ defined on the discrete space $\boldsymbol{\sigma} \in \{-1,1\}^N$, where $N$ is the number of variables or spins.

First, discrete variables $\boldsymbol{\sigma}$ are relaxed to the continuous space $\boldsymbol{x} \in \mathbb{R}^N$ and we define the potential (or energy function) $V(\boldsymbol{x})$ on the real space. The following dynamical system is used to direct the search to lower energy states:
\begin{gather}
\gamma \frac{d^2 \boldsymbol{u}}{dt^2} + \frac{d \boldsymbol{u}}{dt} = - \alpha \boldsymbol{u} - \boldsymbol{e} \circ \nabla_x V,\label{eq:1}
\end{gather}
\noindent where the symbol $\circ$ denote the hadamard product; $\nabla_x V$, the gradient of $V$ with respect to the vector $\boldsymbol{x}$; $\alpha$, a positive parameter; and $\boldsymbol{e}$, a vector of positive auxiliary variables. The term proportional to $\gamma$ represent the momentum which has been utilized for non-convex optimization~\citep{Kalinin2023} and sampling~\citep{neal2011mcmc,Xifara2014}. Moreover, we set:
\begin{align}
x_i = \phi(\tilde{\beta} u_i), \forall i \in \{1, \cdots, N\},\label{eq:3}
\end{align}
and $\phi(x) = \frac{2}{1+e^{-x}} - 1$ is a sigmoidal function normalized to the domain $x \in [-1,1]$. 

The auxiliary variables $e_i$ undergo dynamic changes over time, with the goal of rectifying amplitude heterogeneity are given as follows~\citep{Leleu2019}:
\begin{align}
\frac{d \boldsymbol{e}}{dt} = -\xi (\boldsymbol{x} \circ \boldsymbol{x}- a) \circ \boldsymbol{e},\label{eq:2}
\end{align}
\noindent with $a$ and $\xi$ being parameters: $a$ represents target amplitude of the $x$ variables and $\xi$ represents the speed of the error correction dynamics. Because this algorithm adds momentum dynamics to the original chaotic amplitude control equations~\citep{Leleu2019,Leleu2021}, we call it chaotic amplitude control with momentum (CACm). 

In the Appendix section S1, the role of the auxiliary variables $\boldsymbol{e}$ is analyzed and we show that their modulation can be interpreted as a dynamic pre-conditioning of the Hessian of $V$ at the proximity of the discrete state. Because the auxiliary variables $\boldsymbol{e}$ remain positive, they do not change the inertia (number of positive, negative, and zero eigenvalues) of the Hessian of $V$ but can change its directions and eigenvalues. Thus, CAC performs some dynamic deformation of the potential function $V$ which affects its ability to sample local minima determined by the shape of $V$ when compared to other algorithms, as it will be shown in the following.

\subsection{Probabilistic discrete time model and MCMC}

In order to utilize this dynamical system algorithmically, we discretize eqs.~\ref{eq:1}-\ref{eq:3} using a straightforward Euler approximation. Unlike Hamiltonian Monte Carlo that requires a careful selection of the integration method, an accurate integration scheme is not necessarily needed in this context, as the system can be shown to be bounded due to the influence of the sigmoid function and is dissipative. The utilization of a simple integration method and large integration time step enables the discrete-time approximation of the system to function as computationally efficient solver for combinatorial optimization. The discrete-time model is characterized as follows (after a change of variables):
\begin{align}
\boldsymbol{u}(t+1) &= \boldsymbol{u}(t) - \alpha \boldsymbol{u}(t) - \boldsymbol{e} \nabla_x V (\boldsymbol{x}(t)) + \gamma(\boldsymbol{u}(t)-\boldsymbol{u}(t-1)) ,\label{eq:4}\\
\boldsymbol{e}(t+1) &= \boldsymbol{e}(t)  - \xi (\boldsymbol{x}(t) \circ \boldsymbol{x}(t) - a) \circ \boldsymbol{e}(t).\label{eq:5}
\end{align}
To maintain the effect of the deterministic dynamics when used for optimization, we formulate an MCMC as follows. The eqs.~\ref{eq:4} and \ref{eq:5} are iterated for $n$ time steps, subsequent to which we sample a discrete state $\boldsymbol{\sigma} \in \{-1,1\}^N$ as detailed below:
\begin{align}
P(\sigma_i=1)=\frac{x_i+1}{2}, \sigma_i=-1 \text{ otherwise.} \label{eq:6}
\end{align}
It is worth noting that all spins are updated synchronously and independently. While the deterministic dynamics over $n$ time-steps can be perceived as local updates, the generation of $\boldsymbol{\sigma}$ can leap to remote states. Note moreover that when we set $\xi=0$, $e_i(0)=0$, $\alpha=1$, $t=1$, and update one spin $i$ at a time, we revert to the equation of the Boltzmann machine.

\subsection{Metropolis adjusted chaotic sampling}

When $\xi \neq 0$, detailed balance is violated, causing the stationary distribution of the chain to differ from the Boltzmann distribution. To design an MCMC method that converges (at least theoretically) to the Boltzmann distribution, we introduce a Metropolis-Hastings acceptance criterion for the discrete state samples $\sigma(kt)$, where $k$ belongs to $\{1, 2, ..., K\}$. The inverse of the number of Metropolis-Hastings steps is denoted $f_{\text{MH}} = \frac{1}{K}$ and the total number of steps is noted $T$ with $T=K n$.

Typically, the acceptance criterion of a new configuration $\boldsymbol{\sigma}^2$, when the current state is $\boldsymbol{\sigma}^1$ is given as follows:
\begin{align}
A(\boldsymbol{\sigma}^2/\boldsymbol{\sigma}^1) = \text{min}(1,\frac{P(\boldsymbol{\sigma}^2) Q(\boldsymbol{\sigma}^1/ \boldsymbol{\sigma}^2)}{P(\boldsymbol{\sigma}^1) Q(\boldsymbol{\sigma}^2 / \boldsymbol{\sigma}^1)}). \label{eq:8}
\end{align}
\noindent Here, $Q(\boldsymbol{\sigma}^2/ \boldsymbol{\sigma}^1)$ represents the probability of transitioning from $\boldsymbol{\sigma}^1$ to $\boldsymbol{\sigma}^2$. The Monte Carlo step for our method is comprised of two substeps as follows: (1) A deterministic trajectory following eqs.~\ref{eq:4} and \ref{eq:5} for $n$ steps starting from $\boldsymbol{u}(kn)=\vec{0}$, $\boldsymbol{e}(kn)=\vec{0}$, and , $\boldsymbol{x}(kn) = \boldsymbol{\sigma}^1(kn)$, (2) A probabilitic sample $\boldsymbol{\sigma}^2$ generated with the probability $P^2$ with $P^2 = \frac{\boldsymbol{x}^2+1}{2}$ with $\boldsymbol{x}^2$ defined as $\boldsymbol{x}^2 = \boldsymbol{x}((k+1)n)$.

It is crucial to note that $\boldsymbol{x}((k+1)n)$ is inherently reliant on $\boldsymbol{\sigma}^1$ due to the fact that it is generated from the path of the deterministic dynamical system which commences from $\boldsymbol{\sigma}^1$. Because of the synchronous updating of all spins, we can write the following definition for $Q(\sigma^2/ \sigma^1)$:
\begin{align}
\text{log } Q(\boldsymbol{\sigma}^2/ \boldsymbol{\sigma}^1) = \sum_{i=1}^N  
 \left[\frac{1+\boldsymbol{\sigma}^2_i}{2} \text{log } (P_i^2) +  \frac{1-\boldsymbol{\sigma}^2_i}{2} \text{log } (1-P_i^2) \right]. \label{eq:9}
\end{align}
To calculate the transition probability in the reverse Monte Carlo step from $\boldsymbol{\sigma}^2$ to $\boldsymbol{\sigma}^1$, we need to apply first the deterministic sub-step (1) from $\boldsymbol{\sigma}^2$ to obtain a candidate end point $\tilde{\boldsymbol{x}}^1$, from which we can calculate the probability to generate the random sample $\boldsymbol{\sigma}^1$ with probability $\tilde{P}^1 = \frac{\tilde{\boldsymbol{x}}^1+1}{2}$. The acceptance criterion can then be rewritten as follows:
\begin{align}
\text{log } A(\boldsymbol{\sigma}^2/\boldsymbol{\sigma}^1) &= \text{min}(0, \beta (V(\boldsymbol{\sigma}^1)-V(\boldsymbol{\sigma}^2) - \sum_{i=1}^N  \left[\frac{1+\boldsymbol{\sigma}^2_i}{2} \text{log } (P_i^2) +  \frac{1-\boldsymbol{\sigma}^2_i}{2} \text{log} (1-P_i^2) \right] \nonumber \\ 
& + \sum_{i=1}^N  \left[ \frac{1+\boldsymbol{\sigma}^1_i}{2} \text{log} (\tilde{P}_i^1) +  \frac{1-\boldsymbol{\sigma}^1_i}{2} \text{log } (1-\tilde{P}_i^1)\right] ).\label{eq:10} 
\end{align}
This acceptance criterion can be computed efficiently since it only requires the logarithm of terms already computed within the deterministic equations. The discrete state $\boldsymbol{\sigma}^2$  is accepted with probability $ A(\boldsymbol{\sigma}^2/\boldsymbol{\sigma}^1) $. The Markov chain using the Metropolis-Hastings step respects detailed balance condition and converges to our target Boltzmann distribution $P(x) = e^{-\beta V(x)}$.

In summary, the pseudo code of our approach is given in section S2 of the Appendix. Note that $\boldsymbol{x}^2$ is solely determined by $\boldsymbol{\sigma}^1$. The deterministic path to $\boldsymbol{x}^2$ originates from $\boldsymbol{\sigma}^1$ with $\boldsymbol{e}=1$. Consequently, the transition probability from $\boldsymbol{\sigma}^1$ to $\boldsymbol{\sigma}^2$ depends only on $\boldsymbol{\sigma}^1$ (mapped to $\boldsymbol{x}^2$ by the deterministic flow), preserving the Markov property. The updates of all spins are done in parallel independently, resulting in a product of probabilities. This can be reduced to a sum of log probabilities for more stable numerical simulations as shown in eq.~\ref{eq:10}. We call the algorithm defined by eqs.~\ref{eq:4}-\ref{eq:10} Metropolis Hastings Chaotic Amplitude Control with momentum (MHCACm).

\subsection{Summary of algorithms}

A key property of new algorithm proposed in this work is that can be interpreted as a generalization of many existing methods which exist as limiting cases of MHCACm. Some examples are: Simulated annealing~\citep{kirkpatrick1983optimization}, Hopfield neural networks~\citep{Hopfield1985}, analog iterative machines~\citep{Kalinin2023}, chaotic amplitude control~\citep{Leleu2019}, and chaotic amplitude control with momentum. These limiting cases are summarized in Table~\ref{tab:1}. The generalized algorithms tend to have more parameters which need to be accurately tuned in order to adequately utilize the dynamics provided by each aspect of the algorithm. When properly tuned to a given class of instances, the performance of the generalized algorithm is at least as good as all of its constituting parts. Importantly, automatic parameter tuning methods have recently been developed which allow the many parameters to be quickly and accurately tuned ~\citep{Reifenstein2024}.

\begin{table}[H]
  \caption{Limit cases of the proposed algorithm MHCACm\label{tab:1}}
  \centering
  \begin{tabular}{l|l|l|l||l|l|l|l}
    \toprule
    model & $f_{\text{MH}}$ & $\gamma$ & $\xi$ & model & $f_{\text{MH}}$ & $\gamma$ & $\xi$ \\
    \midrule
    SA & $=\frac{1}{T}$ & $=0$ & $=0 $ & CAC & $=1$ & $=0$ & $>0$ \\
    HNN & $=1$ & $=0$ & $=0$ & CACm & $=1$ & $>0$ & $>0$ \\
    AIM & $=1$ & $>0$ & $=0$ & MHCACm & $\frac{1}{T}<f_{\text{MH}}<1$ & $>0$ & $>0$ \\
    \bottomrule
  \end{tabular}
\end{table}

\section{Benchmark}

\subsection{Degenerate planted Wishart instances\label{sec:degenerate}}

We construct a set of planted Ising problem instances that exhibit ground states with different tunable properties. The construction of these instances is inspired by the Wishart planted ensemble (WPE)~\citep{hamze2020wishart}. WPE instances are particularly useful for benchmarking due to their tunable hardness parameter $\alpha_{\text{WPE}}$\footnote{$\alpha_{\text{WPE}}$ is set to 0.8 in this paper.}. We introduce the degenerate Wishart planted ensemble (dWPE) to allow for planting of multiple degenerate ground states. Additionally, we add a tunable "bias", denoted $b$, to the energy landscape which allows us to make some ground states easier/harder to find than others for heuristic algorithms as depicted in Fig.~\ref{fig:1} (b). Thus, we refer to these different ground states as "easy" and "hard" solutions. The details of the construction method is described in the Appendix section S5.

\subsection{Computation time to find easy and hard solutions}

In this section, we evaluate the novel algorithm for finding low energy states of non-convex combinatorial optimization problems. A common metric for evaluating the performance of Ising solvers is the "time to solution" (TTS) which measures the number of steps needed to have 99$\%$ probability of finding any ground state. In case there are $D$ degenerate ground states (i.e., solutions of same energy), there is equal probability of sampling them from the Boltzmann distribution. However, algorithms generally reach a ground state before equilibration to the Boltzmann distribution and, consequently, each degenerate solution can have different probability of being sampled at finite sampling time. We define $\text{TTS}$, $\text{TTS}_{\text{easy}}$, and $\text{TTS}_{\text{hard}}$ as the time to reach any, the easier, and harder to find degenerate solution (of same Ising energy) with 99$\%$ probability as follows:
\begin{align}
\text{TTS} = T\frac{\log(1 - 0.99)}{\log \left(1 - \sum_d P_d \right)}, \text{TTS}_{\text{easy}} = T\frac{\log(1 - 0.99)}{\log \left(1 - \text{min}_d P_d  \right)}, \text{TTS}_{\text{hard}} = T\frac{\log(1 - 0.99)}{\log \left(1 - \text{max}_d P_d  \right)},
\end{align}
\noindent where $P_d$ is the probability to sample the ground state $d$. The algorithms used for comparison are all tuned to near optimal parameters by minimizing the \text{TTS}. We use a state-of-the-art autotuning method called dynamic anisotropic smoothing~\citep{Reifenstein2024} (DAS) which is particularly well fitted to optimizing the parameters of heuristics for combinatorial optimization ~\citep{Reifenstein2024}. The use of a automatic parameter tuning also guarantees a fair comparison between algorithms (see Appendix S4). Moreover, all algorithms are compared using the number of ``steps" to solution, where the algorithmic complexity of a single step is dominated by the matrix-vector multiplication of size $N \times N$\footnote{For SA, a step represents a full sweep of $N$ spin updates which also has complexity $O(N^2)$. The same for PT, but divided by the number of replica.}.

\subsection{Global optimization to any optimal solution}

\begin{figure}[h]
  \centering
\includegraphics[width=1.00\columnwidth]{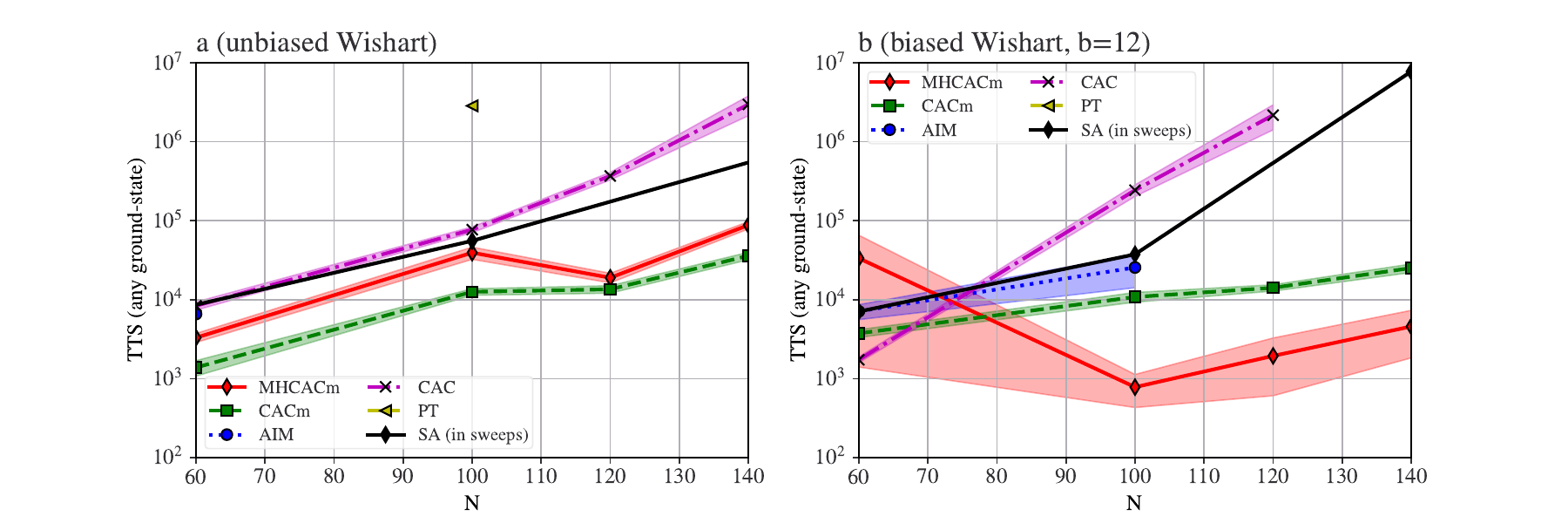}
  \caption{\label{fig:3} The time to solution of (unbiased) Wishart planted instances and biased degenerate planted instance with $b=12$ are shown in (a) and (b), respectively. $f_{\text{MH}} = 0.1$. All other parameters tuned with DAS~\citep{Reifenstein2024} (see Appendix S4). Colored regions represent the 95\% confidence interval calculated by generating 15 instances per data point.}
\end{figure}

For the task of finding any ground state, Fig.~\ref{fig:3} (a) shows that MHCACm performs almost as well as CACm and better than AIM, CAC, and SA\footnote{Results for SA and PT are collected using dwave-neal~\citep{dwave-neal} and pySA~\citep{pySA}}. In the case of sampling any ground state of biased ($b=12$ ) dWPE problem instances (see section \ref{sec:degenerate}), MHCACm requires significantly less steps than other algorithm to find an optimal solution (see Fig. \ref{fig:3} (b)). 

\subsection{Sampling in the case of biased degenerate optimal solutions }

\begin{figure}[H]
  \centering
\includegraphics[width=1.00\columnwidth]{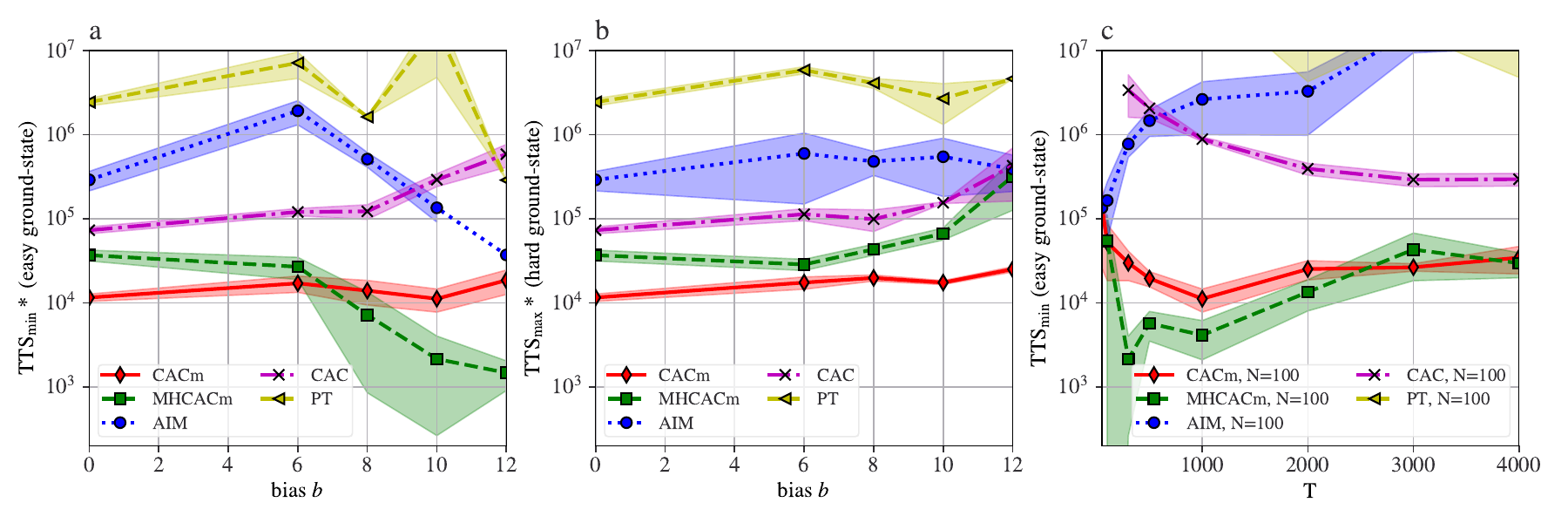}
  \caption{\label{fig:4} Time to sample the easy solution $\text{TTS}_{\text{easy}}$ (a) and hard solution $\text{TTS}_{\text{hard}}$ (b) vs. the bias $b$ of degenerate Wishart planted instances. (c) $\text{TTS}_{\text{easy}}$ vs. runtime $T$ for degenerate Wishart planted instances. The same optimal parameters as in Fig.~\ref{fig:3} are used. (a,b) $N=100$. (c) $N=100$, $b=10$.}
\end{figure}

Next, we examine why MHCACm requires fewer steps to find any solution when there is a larger bias in the dWPE instances. Fig. \ref{fig:4} (a) and (b) show that the $\text{TTS}_{\text{easy}}$ for easy ground states decreases as a function of the bias $b$ of the generated Wishart instances. Conversely, the time to sampled the hard solution $\text{TTS}_{\text{hard}}$ increases. This is unlike the case of CACm without the MH step, for which the $\text{TTS}_{\text{easy}}$ and $\text{TTS}_{\text{hard}}$ remain approximately equal for all bias $b>0$. Thus, MHCACm is better at finding the easier solutions, which explains the smaller overall TTS (to any ground state) observed in Fig.~\ref{fig:3}. Table~\ref{tab:2} indicates that MHCACm is the fastest algorithm to sample from easier ground states, whereas CACm is the better algorithm to reach the harder to find ones. 
\\
\\
This difference in complexity between finding the easy and hard solutions is also present in other MCMC algorithms such as SA and PT as can be seen in Tab.~\ref{tab:2}. The energy landscape is structured so that there are many more low energy states in the vicinity of the easy solution. We also show the results of the Boltzmann sampling algorithm Gibbs with gradient~\citep{Grathwohl2021} (GWG), although its purpose is not combinatorial optimization. An algorithm that attempts to sample from a Boltzmann distribution will thus have a large bias towards this part of the solution space at finite temperature, because of the many low energy states that should be sampled with high probability. Although in the infinite time limit we would expect equal probability of finding all ground states, in practice the extremely slow relaxation time of these algorithms makes it very improbable for the "hard" ground state to be found. However, our hybrid algorithm is able to find both ground states with nonzero probability. We believe a reason for this is that the purely analog methods such as CAC, AIM and CACm do not sample from a Boltzmann distribution and thus can partially avoid these large basins of attraction. We see this in Tab.~\ref{tab:2} in which the analog dynamics are able to find both ground states with reasonable success as well.

\begin{table}[H]
  \caption{Time to solution (in steps, i.e., matrix vector multiplication) using degenerate Wishart planted instances with bias $b=12$. We show both TTS, $\text{TTS}_{\text{easy}}$ and $\text{TTS}_{\text{hard}}$ (as defined in the text) obtained for a variety of algorithms and two problem sizes. Since SA and GWG were not able to sample the hard ground state, $\text{TTS}_{\text{hard}}$ is not defined for it and $\text{TTS}_{\text{hard}}$ = TTS. Bold numbers denote the best algorithm.
  \label{tab:2}}
  \centering
  \begin{tabular}{l||l|l||l|l||l|l}
    \toprule
    {} & \multicolumn{2}{|c||}{$10^{-3}$ TTS} & \multicolumn{2}{|c||}{$10^{-3}$ $\text{TTS}_{\text{easy}}$}& \multicolumn{2}{|c}{$10^{-3}$ 
 $\text{TTS}_{\text{hard}}$}\\
    \midrule
    model/N & $100$ & $140$ & $100$ & $140$ & $100$ & $140$ \\
    \midrule
    SA & 27.56 & 1917.67 & 27.56 & 1917.67 & N/A & N/A \\
    PT & 2437.92 & N/A & 287.71 & N/A & 4605.06 & N/A  \\
    \midrule
    GWG & 6369.48 & 15779.94 & 6369.48 & 15779.94 & N/A & N/A \\
    \midrule
    AIM &  22.31 &  906.96 & 37.06&958.13&390.72 &16884.93 \\
    CAC &  238.81 &  5833.26 & 586.26&46960.90&427.34 &6634.85 \\
    CACm &  9.41 &  22.91 & 18.54&37.84&\textbf{25.08} &\textbf{76.88} \\
    \midrule 
    MHCACm &  \textbf{1.40} &  \textbf{2.44} & \textbf{1.47}&\textbf{2.44}&318.16 &248.63 \\
    \bottomrule
  \end{tabular}
\end{table}

\subsection{Results on GSET}

The GSET ~\citep{GSET} is a set of Max-Cut problem instance which are commonly used to benchmark Ising and Max-Cut solvers ~\citep{goto2021high, Leleu2019, BenlicBLS}. To test the ability of MHCACm to find optimal solutions on larger optimization problems, we briefly present some results on these instances in comparison with a state of the art algorithm known as dSBM~\citep{goto2021high}. Table~\ref{tab:4} shows that MHCACm reaches best known solutions faster than dSBM~\citep{goto2021high} on the first few instances of the GSET problem set~\citep{GSET}.

\begin{table}[H]
  \caption{TTS (in steps or MVM) for solving GSET~\citep{GSET} instances of MHCACm and dSBM~\citep{goto2021high}. \label{tab:4}}
  \centering
  \begin{tabular}{l||l|l|l||l|l|l}
    \toprule
    instance & \multicolumn{3}{c||}{MHCACm} & \multicolumn{3}{|c}{dSBM}\\ 
    \midrule
    $N$ (id) & $p_0$  &  T  & TTS & $p_0$  &  T  & TTS \\
    \midrule
    800 (1) & 0.612000 & 3000 & \textbf{14592} & 0.026777 & 2000 & 339332 \\
    800 (2) & 0.021000 & 3000 & \textbf{650949} & 0.010660 & 6000 & 2578124 \\
    800 (3) & 0.533500 & 3000 & \textbf{18118} & 0.033920 & 3000 & 400343 \\
    800 (4) & 0.140500 & 3000 & \textbf{91249} & 0.025144 & 2000 & 361673 \\
    800 (5) & 0.044000 & 3000 & \textbf{307029} & 0.022099 & 3000 & 618221 \\
    800 (6) & 0.070500 & 3000 & \textbf{188972} & 0.023856 & 1000 & 190728 \\
    800 (7) & 0.300000 & 3000 & \textbf{38734} & 0.022552 & 1000 & 201889 \\
    800 (8) & 0.166500 & 3000 & \textbf{75858} & 0.019060 & 1500 & 358947 \\
    \bottomrule
  \end{tabular}
\end{table}


\subsection{Parallelization on GPU}

The algorithm is particularly well-suited for deployment on highly parallel hardware, such as GPUs, due to its capacity for parallel spin updates. In particular, the computational bottleneck of MHCACm is the matrix vector multiplication (MVM), a basic linear algebra operation that can be done very efficiently on modern GPUs using frameworks such as PyTorch. In table~\ref{tab:3} we compare the wall-clock TTS of our PyTorch implementation of MHCACm finding ground states of a degenerate WPE instance of size $N=100$. We also show the wall-clock time of a standard implementation of SA ~\citep{qubovert} running an CPU. MHCACm on GPU exhibits a 100x speed up in real time against simulated annealing on CPU for sampling easy optimal solutions. Furthermore, MHCACm is compatible with an implementation on special-purpose analog hardware designed for linear algebra acceleration, such as a memresistor crossbar~\citep{hu2014memristor} or on-chip photonic mesh arrays~\citep{bogaerts2020programmable}.

\begin{table}[H]
  \caption{Wall clock times of algorithms on different computing platforms. TTS is time to find any ground state of a degenerate WPE instance with $\alpha_{\text{WPE}} = 0.8, N=100$ and bias $b$ = 12. The GPU used is a Nvidia V100 and the CPU wall time is calculated on a 2024 MacBook Pro (Apple M3 Chip). \label{tab:3}}
  \centering
  \begin{tabular}{l||l|l}
    \toprule
      {} & \multicolumn{2}{c}{TTS (s) ($N=100$)}\\ 
      \midrule
      &  unbiased WPE  & biased WPE ($b=12$)  \\
    \midrule
    SA (CPU) & $5.794 \times 10^{-7}$ & $0.0217$ \\
    MHCACm (CPU) & $1.739 \times 10^{-6}$ & $0.00243$ \\
    MHCACm (GPU) & $\mathbf{9.798 \times 10^{-8}}$  & $\mathbf{1.373 \times 10^{-4}}$ \\
    \bottomrule
  \end{tabular}
\end{table}

\section{Comparison to other approaches}

Methods such as HMC~\citep{neal2011mcmc,Xifara2014} introduce auxiliary degrees of freedom (momentum) to the gradient dynamics, providing the system additional directions to move away from local minima of the error function. However, these methods do not force the dynamics to be close to the discrete state, rendering them ill-suited for discrete state sampling. CAC suggests a different mechanism for the auxiliary variables. It adjusts the amplitude of the continuous variables, ensuring proximity to the discrete (binary) state~\citep{Leleu2019,Leleu2021}. Conventionally, this can be interpreted as a dual-primal Lagrangian problem in which we want to minimize the error function (in the continuous embedding) while also satisfying a constraint (being close to the discrete state). Our method differs from the primal dual~\citep{Vadlamani2020} approach. With CAC, the error function is achieved by multiplying the gradient with auxiliary variables, akin to dynamically pre-conditioning the Hessian of the cost function.

Unlike HMC however, CAC does not conserve (even approximately) energy. In HMC~\citep{neal2011mcmc,Xifara2014}, the MH step compensates for the discretization errors brought about by the leapfrog integration. However, the reason for adding an MH step to CAC is different. State-of-the-art algorithms that exploit relaxation to an continuous state form discrete combinatorial optimization such as memcomputing~\citep{sheldon2019taming} chaotic amplitude control~\citep{Leleu2019}, and simulated bifurcation machine~\citep{goto2021high} do not sample fairly from the Boltzmann distribution. Our proposed methodology bridges this gap, integrating continuous relaxation algorithms with fair sampling of a discrete distribution via the construction of a hybrid MCMC.

\section{Conclusion}

The MHCACm algorithm effectively combines chaotic dynamics with the Metropolis-Hastings method to enhance ground-state sampling efficiency in discrete spaces, ensuring fair sampling at equilibrium while biasing towards more reachable ground states during non-equilibrium dynamics. This approach significantly improves the speed and accuracy of finding optimal solutions for combinatorial optimization tasks, demonstrating superior performance and suitability for large-scale parallel processing on GPUs.


The theoretical understanding of the sampling capabilities of MHCACm are out of the scope of this paper. However, numerical evidence seems to indicate that the algorithm not only achieves fair sampling from the Boltzmann distribution after sufficient relaxation time (see Apendix Fig.~S2) but can exhibit faster relaxation times on NP-hard Ising problem instances with complicated energy landscape (see Fig.~\ref{fig:4} and Tab.~\ref{tab:1}) when compared to traditional methods such as SA. Lastly, the connection with learning in over-parameterized neural networks is to be investigated. Finding the most reachable and common solution to a large scale optimization problem is particularly useful for training neural networks in machine learning, as it typically corresponds to a trained network with better generalization properties~\citep{Soudry2018,baity2018comparing,feng2021inverse,baldassi2022learning,baldassi2023typical}.

\clearpage


\bibliographystyle{iclr2025_conference}

\clearpage

\section*{- Appendix -\\ Non-Equilibrium Dynamics of Hybrid Continuous-Discrete Ground-State Sampling}

\beginsupplement 

\section{Analysis of the deterministic path}

In the following section, we analyze the deterministic dynamical system defined as chaotic amplitude control with momentum (CACm) in the main manuscript. For simplicity, we do not take into account the effect of momentum and set $\gamma=0$.

\subsection{Dynamic pre-conditioning of the Hessian of $V$}

The equation (1) of the main manuscript is a straightforward gradient descent of the potential $V$. However, the gradient is modulated by the auxiliary variables. In this section, we give more context about the motivation for using the auxiliary variables $\boldsymbol{e}$ and explain that their modulation can be interpreted as a dynamic pre-conditioning of the Hessian of $V$ at the proximity of the discrete state $\boldsymbol{\sigma}$. First, we show in Appendix section \ref{sec:changevar} that the variables $\boldsymbol{u}$ in the system of eqs. (1) and (2) can be eliminated by a change of variables. Then, the Jacobian matrix $J_{xx}$ in the space $\boldsymbol{x}$ defined as $J_{xx} = \{\frac{\partial \dot{x}_i}{\partial x_j}\}_{ij}$ with $\dot{x}_i = \frac{d\boldsymbol{x}}{dt}$ can be expressed as follows at proximity of the fixed points of the system (see Appendix section \ref{sec:linearstab}):
\begin{align}
J_{xx} = D[-\alpha f(\sqrt{a} \boldsymbol{\sigma})] - D[\boldsymbol{e} \circ g(\sqrt{a} \boldsymbol{\sigma})] - D[\boldsymbol{e} h(\sqrt{a} \boldsymbol{\sigma})] H(\sqrt{a}\boldsymbol{\sigma}) \label{eq:S2b}
\end{align}

\noindent where $D[\boldsymbol{x}]$ is a diagonal matrix with elements given by the components of the vector $\boldsymbol{x}$ and $H(\sqrt{a}\boldsymbol{\sigma})$ is the Hessian of $V$ defined as the Jacobian of $\nabla_x V$ and calculated at the point $\sqrt{a}\boldsymbol{\sigma}$. The functions $f$ and $g$ and defined in Appendix section \ref{sec:changevar}. Equation \ref{eq:S2b} shows that the auxiliary variables $\boldsymbol{e}$ modulates the Hessian $H$ on the slow time scale such that $\tilde{H}(t,\tau) = D[e(\tau) h(x_i)] H$ with $h(x) = \frac{1}{\frac{d \psi}{d x}} > 0$ for $x \in ]-1,1[$ and $\psi (x) = \frac{\phi^{-1}(x)}{\tilde{\beta}}$ (see lemma 1 in Appendix section \ref{sec:linearstab}).

Given that the fixed points of eqs. (1) and (2) are given as $e_i = - \alpha \sqrt{a} \frac{1}{\sigma_i \frac{\partial V}{\partial x_i} ( \sqrt{a}\sigma_i)}$, we can interpret the role of $\boldsymbol{e}$ as a normalization of each row of the Hessian by the inverse of the gradient in the corresponding direction  (see theorem 1 in Appendix section \ref{sec:linearstab}). Because the auxiliary variables $\boldsymbol{e}$ remain positive, they do not change the inertia (number of positive, negative, and zero eigenvalues) of the Hessian $H$ (see lemma 2 in Appendix section \ref{sec:effectiveH}) but can change its directions and eigenvalues. Moreover, we show in the theorem 2 of in Appendix section \ref{sec:eigen} that this dynamical system can exhibit period cycles and chaos when the target amplitude $a$ is chosen to be sufficiently small. In summary, the role of the auxiliary variables is to act as a pre-conditioning of the Hessian and introduce entropy in the search dynamics due to its chaotic dynamics. Similar dynamics have been shown to be competitive with state-of-the-art heuristics to solve Ising problems~\citep{Leleu2019,Leleu2021}.

\subsection{Change of variable\label{sec:changevar}}

Consider $x_i = \phi(2 \beta u_i) \implies u_i = \psi_{\beta} (x_i)$ with $\psi_{\beta} (x_i) = \frac{\phi^{-1}(x_i)}{2 \beta}$.

In the case $\phi(x) = \frac{2}{1+e^{-x}} - 1$, we have $\phi^{-1}(x) = -ln(\frac{2}{x + 1} - 1)$, $\psi_{\beta}'(x) = -\frac{1}{\beta (x + 1) (x - 1)}$ with $\psi_{\beta}'(x) = \frac{ \partial \psi_{\beta}(x)}{\partial x}$. Moreover, $\psi_{\beta}' (x)$ is positive on $]-1,1[$ and $\frac{1}{\psi_{\beta}' (x)} \rightarrow 0$ for $|x|\rightarrow0$.

Because $\frac{d u_i}{dt} = \psi_{\beta}' (x_i) \frac{d x_i}{dt}$ with $\psi_{\beta}' = \frac{1}{\beta} \frac{\partial \phi^{-1}(x_i)}{\partial x_i}$, the equations of motion can be rewritten by eliminating $u_i$ ($\forall i \in \{1, \cdots, N\}$) as follows:

\begin{align}
\frac{d x_i}{dt} &= - \alpha \frac{\psi_{\beta} (x_i)}{\psi_{\beta}' (x_i) } - \frac{e_i}{\psi_{\beta}' (x_i) } \nabla_x V,\label{eq:S1}\\
\frac{d e_i}{dt} &= -\xi (x_i^2 - a) e_i\label{eq:S2},
\end{align}

\subsection{Jacobian matrix at the fixed points\label{sec:linearstab}}

The fixed points $\frac{d x_i}{dt}=0$ and $\frac{d e_i}{dt}=0$ of the dynamical system described by eqs.~\ref{eq:S1} and \ref{eq:S2} are given as follows:

\begin{align}
x_i &= \sigma_i \sqrt{a},\label{eq:S3} \\
e_i &= - \frac{\alpha \psi_{\beta} (\sigma_i \sqrt{a})}{\frac{\partial V}{\partial x_i}(\sigma \sqrt{a})}.\label{eq:S4}
\end{align}

Moreover, the Jacobian matrix $J$ at the fixed points is defined as:

\begin{equation}
J = \left[ \begin{array}{cc}
J_{xx} & J_{xe} \\
J_{ex} & J_{ee} 
\end{array} \right]
\end{equation}

\noindent with

\begin{align}
J_{xx} &= D[-\alpha f(\sqrt{a} \boldsymbol{\sigma})] - D[\boldsymbol{e} \circ g(\sqrt{a} \boldsymbol{\sigma})] - D[\boldsymbol{e} \circ h(\sqrt{a} \boldsymbol{\sigma})] H(\sqrt{a} \boldsymbol{\sigma}),\label{eq:S5}\\
J_{xe} &= - D[\{\frac{\frac{\partial V}{\partial x_i}(\sqrt{a} \sigma_i)}{\psi_{\beta}' (\sqrt{a}\sigma_i) }\}_i],\label{eq:S6}\\
J_{ex} &= -2 \xi \sqrt{a} D[\boldsymbol{\sigma} \circ \boldsymbol{e}],\label{eq:S7}\\
J_{ee} &= 0,\label{eq:S8}
\end{align}

\noindent with $H(x)=J[\nabla_x V]$ is the Hessian of $V$ in $x$ without its diagonal elements. The notation $D[\boldsymbol{x}]$ denote the diagonal matrix with diagonal elements given by the components of the vector $\boldsymbol{x}$. The functions $f$, $g$, and $h$ of eq.~\ref{eq:S5} are defined as follows:

\begin{align}
f(x_i) &= \frac{\partial \frac{\psi_{\beta}}{\psi_{\beta}' (x_i)}}{\partial x_i},\label{eq:S9}\\
g(x_i) &= \frac{\partial \frac{1}{\psi_{\beta}' (x_i)} \frac{\partial V}{\partial x_i}}{\partial x_i},\label{eq:S10}\\
h(x_i) &= \frac{1}{\psi_{\beta}' (x_i)} > 0\text{ for }x_i \in ]-1,1[.\label{eq:S11}
\end{align}

In the case of $\phi(x) = \frac{2}{1+e^{-x}}-1$, we have:

\begin{align}
f(x_i) &= 2 + x_i \text{ln}(-\frac{-1+x_i}{1+x_i}),\\
g(x_i) &= - \beta x_i \frac{\partial V}{\partial x_i} - \beta (x_i + 1) (x_i - 1) \frac{\partial^2 V}{\partial x_i^2},\\
h(x_i) &= - \beta (x_i + 1) (x_i - 1).
\end{align}

Note that $f$ is an even function with $f(x)=f(-x)$.

In the case $\xi \rightarrow 0$, the dynamics of $e_i$ variables becomes much slower than that of $x_i$. We note $\tau$ the slower time scale of $e_i$ with $\tau = t \xi$. The lemma 1 is obtained using eq.~\ref{eq:S5}.

\textbf{Lemma 1}: When $\xi \rightarrow 0$, the Hessian $H$ of $V$ is modulated on a slow time scale $\tau$ with $\tilde{H} = D[\boldsymbol{e} \circ g(\sqrt{a} \boldsymbol{\sigma})] H$ as proximity of the fixed points $\boldsymbol{x} = \sqrt{a} \boldsymbol{\sigma}$. We call $\tilde{H}$ the effective Hessian matrix.

\subsection{Effective Hessian\label{sec:effectiveH}}

Stability of the fixed points is determined by the eigenvalues of the Jacobian matrix $J$. The blocks $J_{ex}$ and $J_{xe}$ are such that (see also~\citep{Leleu2019}):

\begin{align}
J_{ex}J_{xe} = b I, \label{eq:S12}
\end{align}

\noindent with $I$ the identity matrix and $b= -2  \xi \alpha \sqrt{a} \frac{\psi_{\beta} (\sqrt{a})}{\psi_{\beta}' (\sqrt{a})}$ and $J_{ee}=0$. Consequently, the determinant $P(\lambda)=|J-\lambda I|$ characterizing the eigenvalues of $J$ has the following property:

\begin{align}
P(\lambda) = \left| \begin{array}{cc}
J_{xx} - \lambda I & J_{xe} \\
J_{ex} & - \lambda I
\end{array} \right| = |J_{xx} - \lambda I| \lambda - J_{xe}J_{ex} = |J_{xx} - \lambda I| \lambda - bI. \label{eq:S13}
\end{align}

If we note $\mu_i$ the eigenvalues of $J_{xx}$, the characteristic polynomial can be rewritten:

\begin{align}
P(\lambda)  = -(\mu - \lambda) \lambda - bI = \lambda^2 - \mu \lambda - b. \label{eq:S14}
\end{align}

\noindent The eigenvalues $\lambda_j$ can come in pairs $\lambda_j^{+}$ and $\lambda_j^{-}$ depending on the sign of $\Delta_j = \mu_j^2 +4b$ and are expressed as follows by solving $P(\lambda)=0$  ($\forall j \in \{1, \cdots, N\}$):

\begin{align}
\lambda_j^{\pm} = (\mu_j \pm \sqrt{\Delta_j})\frac{1}{2} \text{ if } \Delta_j>0,\label{eq:S15}\\
\lambda_j^{\pm} = (\mu_j \pm i \sqrt{\Delta_j})\frac{1}{2} \text{ if } \Delta_j<0.\label{eq:S16}
\end{align}

We define the pre-conditioned Hessian matrix $\tilde{\tilde{H}}$ as $\tilde{\tilde{H}} ( \sqrt{a} \boldsymbol{\sigma}) = D[\frac{\sigma_i}{\frac{\partial V}{\partial x_i}(\sigma_i \sqrt{a})}] H(\sqrt{a}\sigma_i)$ such that :

\begin{align}
D[e_i h(\sqrt{a}\sigma_i)] H(\sqrt{a}\boldsymbol{\sigma}) &= D[-\alpha \frac{1}{\frac{\partial V}{\partial x_i}(\sigma_i \sqrt{a})} \frac{ \psi_{\beta} (\sigma_i \sqrt{a})}{\psi_{\beta}'(\sigma_i \sqrt{a})}] H(\sqrt{a}\boldsymbol{\sigma}),\\
&= -\alpha  \frac{ \psi_{\beta} (\sqrt{a})}{\psi_{\beta}'(\sqrt{a})} D[\frac{\sigma_i}{\frac{\partial V}{\partial x_i}(\sigma_i \sqrt{a})}] H(\sqrt{a}\boldsymbol{\sigma}),\\
&= -\alpha  \frac{ \psi_{\beta} (\sqrt{a})}{\psi_{\beta}'(\sqrt{a})} \tilde{\tilde{H}} (\sqrt{a} \boldsymbol{\sigma}),
\end{align}

\noindent The last step uses the fact that $\frac{ \psi_{\beta} (x)}{\psi_{\beta}'(x)}$ is an odd function. 

Note that the factors $\frac{\sigma_i}{\frac{\partial V}{\partial x_i}(\sigma_i \sqrt{a})}$, $\forall i$, multiplying the Hessian $H$ are positive at local minima satisfying $\frac{\partial V}{\partial x_i}(\sigma_i \sqrt{a}) \sigma_i >0$, $\forall i$. In this case, the eigenvalues of $D[\frac{\sigma_i}{\frac{\partial V}{\partial x_i}(\sigma_i \sqrt{a})}] H(\sqrt{a}\sigma_i)$ are the same as $D[\frac{\sigma_i}{\frac{\partial V}{\partial x_i}(\sigma_i \sqrt{a})}]^{1/2} H(\sqrt{a}\sigma_i) D[\frac{\sigma_i}{\frac{\partial V}{\partial x_i}(\sigma_i \sqrt{a})}]^{1/2}$. By Sylvester's law of matrix inertia, this implies that the eigenvalues of $\tilde{\tilde{H}}$ and $H$ have the same signs. In matrix theory, the term ''inertia" refers to the number of positive, negative, and zero eigenvalues of a real symmetric matrix. 

\textbf{Lemma 2}: The signs of the eigenvalues of $\tilde{\tilde{H}}$ and $H$ are the same when $\sigma_i \frac{\partial V}{\partial x_i}(\sigma_i \sqrt{a}) > 0$, $\forall i$.

\textbf{Theorem 1}: The effect of error variables at proximity of fixed points is to replace the Hessian $H$ of $V$ by an effective Hessian $\tilde{\tilde{H}}$ with $\tilde{\tilde{H}} ( \sqrt{a} \boldsymbol{\sigma}) = D[\frac{\sigma_i}{\frac{\partial V}{\partial x_i}(\sigma_i \sqrt{a})}] H(\sqrt{a}\sigma_i)$. The inertia of the Hessian and effective Hessian are the same for local minima of $V$ verifying the condition $\sigma_i \frac{\partial V}{\partial x_i}(\sigma_i \sqrt{a}) > 0$, $\forall i$.

\subsection{Eigenvalues of $H$ and stability of the fixed points\label{sec:eigen}}

The eigenvalues of the Jacobian $J_{xx}$, noted $\mu_j$, can be expressed using the eigen spectrum of the pre-conditioned Hessian $\tilde{\tilde{H}}$, noted $\gamma_j$, as follows using eq.~\ref{eq:S5}:

\begin{align}
\mu_j = -\alpha f(\sqrt{a} \sigma_i) - e_i g(\sqrt{a}\sigma_i) + \alpha  \frac{ \psi_{\beta} (\sqrt{a})}{\psi_{\beta}'(\sqrt{a})} \gamma_j. \label{eq:S17}
\end{align}

The fixed points become stable when the real parts of eigenvalues are all negative, which is verified when the eigenvalue with maximal real part becomes equal to 0 given by the following condition using eqs.~\ref{eq:S15} and \ref{eq:S16}:

\begin{align}
\text{max}_j (\text{Re}[\mu_j \pm \sqrt{\mu_j^2 +4b}]) = 0 \text{ if } \mu_j^2 -4b>0,\label{eq:S18}\\
\text{max}_j  (\text{Re}[\mu_j \pm i \sqrt{-\mu_j^2 -4b}]) = 0 \text{ if } \mu_j^2 -4b<0.\label{eq:S19}
\end{align}

Noting $\tilde{\gamma}$ the eigenvalue with maximal part $\text{Re}[\gamma_j]$, eqs.~\ref{eq:S18} and \ref{eq:S18} imply that the fixed point $\sigma_i \sqrt{a}$ becomes stable under the condition (see eq.~\ref{eq:S16}):

\begin{align}
\tilde{\mu} = 0,\label{eq:S20}\\
\text{i.e., } -\alpha f(\sqrt{a} \sigma_i) - e_i g(\sqrt{a}\sigma_i) - \tilde{\gamma} = 0.\label{eq:S21}
\end{align}

Assuming $\frac{\partial^2 V}{\partial x_i^2} =0$, we have (see eqs.~\ref{eq:S2} and \ref{eq:S10}):

\begin{align}
e_i g(x_i) &=  \frac{\psi_{\beta}'' (x_i)}{(\psi_{\beta}' (x_i))^2} \frac{\partial V}{\partial x_i} \frac{\alpha \psi_{\beta} (x_i)}{\frac{\partial V}{\partial x_i}},\\
&= \alpha  \frac{\psi_{\beta}'' (x_i)}{(\psi_{\beta}' (x_i))^2} \psi_{\beta} (x_i). \label{eq:S22}
\end{align}

Moreover, we have  (see eq.~\ref{eq:S9}):

\begin{align}
f(x_i) &= \frac{\partial \frac{\psi_{\beta}}{\psi_{\beta}' (x_i)}}{\partial x_i}\\
&= 2- \frac{\psi_{\beta}\psi_{\beta}''}{(\psi_{\beta}' (x_i))^2}. \label{eq:S23}
\end{align}

Combining eqs.~\ref{eq:S22} and \ref{eq:S23} , eq. \ref{eq:S21} can be written as follows:

\begin{align}
\tilde{\gamma} (\sqrt{a} \boldsymbol{\sigma}) = 2 \frac{\psi_{\beta}'(\sqrt{a})}{\psi_{\beta} (\sqrt{a})} .\label{eq:S24}
\end{align}

\textbf{Theorem 2}: For small values of the target amplitude $a$ such that $\tilde{\gamma} (\sqrt{a} \boldsymbol{\sigma}) > 2 \frac{\psi_{\beta}'(\sqrt{a})}{\psi_{\beta} (\sqrt{a})}$ where $\tilde{\gamma} (\sqrt{a} \boldsymbol{\sigma})$ is the eigenvalue of  $\tilde{\tilde{H}} ( \sqrt{a} \boldsymbol{\sigma})$ with $\tilde{\tilde{H}} ( \sqrt{a} \boldsymbol{\sigma}) = D[\frac{\sigma_i}{\frac{\partial V}{\partial x_i}(\sigma_i \sqrt{a})}] H(\sqrt{a}\sigma_i)$, the system described by eqs.~\ref{eq:S1} and \ref{eq:S2} does not have any stable fixed points assuming $\frac{\partial^2 V}{\partial x_i^2} =0$.

In general settings where the system is bounded, Theorem 2 implies that the dynamical system of eqs.~\ref{eq:S1} and \ref{eq:S2} settles to either a periodic (quasi periodic) or chaotic attractor for sufficiently large runtime T.

In the case of the Ising Hamiltonian $V(\boldsymbol{x}) = - \frac{1}{2} \sum_{ij} \omega_{ij} x_j x_i$, the calculations are simplified because $\frac{\partial V}{\partial x_i}(\sigma_i \sqrt{a}) = \sqrt{a} \frac{\partial V}{\partial x_i}(\sigma_i)$ and $V(\boldsymbol{\sigma} \sqrt{a}) = a V(\boldsymbol{\sigma})$ and eq. \ref{eq:S24} can be written:

\begin{align}
\tilde{\gamma} (\boldsymbol{\sigma})  = 2 \sqrt{a} \frac{\psi_{\beta}'(\sqrt{a})}{\psi_{\beta} (\sqrt{a})} ,\label{eq:S25}
\end{align}

\noindent where $\tilde{\gamma}$ is independent of $\sqrt{a}$.

\newpage

\begin{figure}
  \centering
\includegraphics[width=1.00\columnwidth]{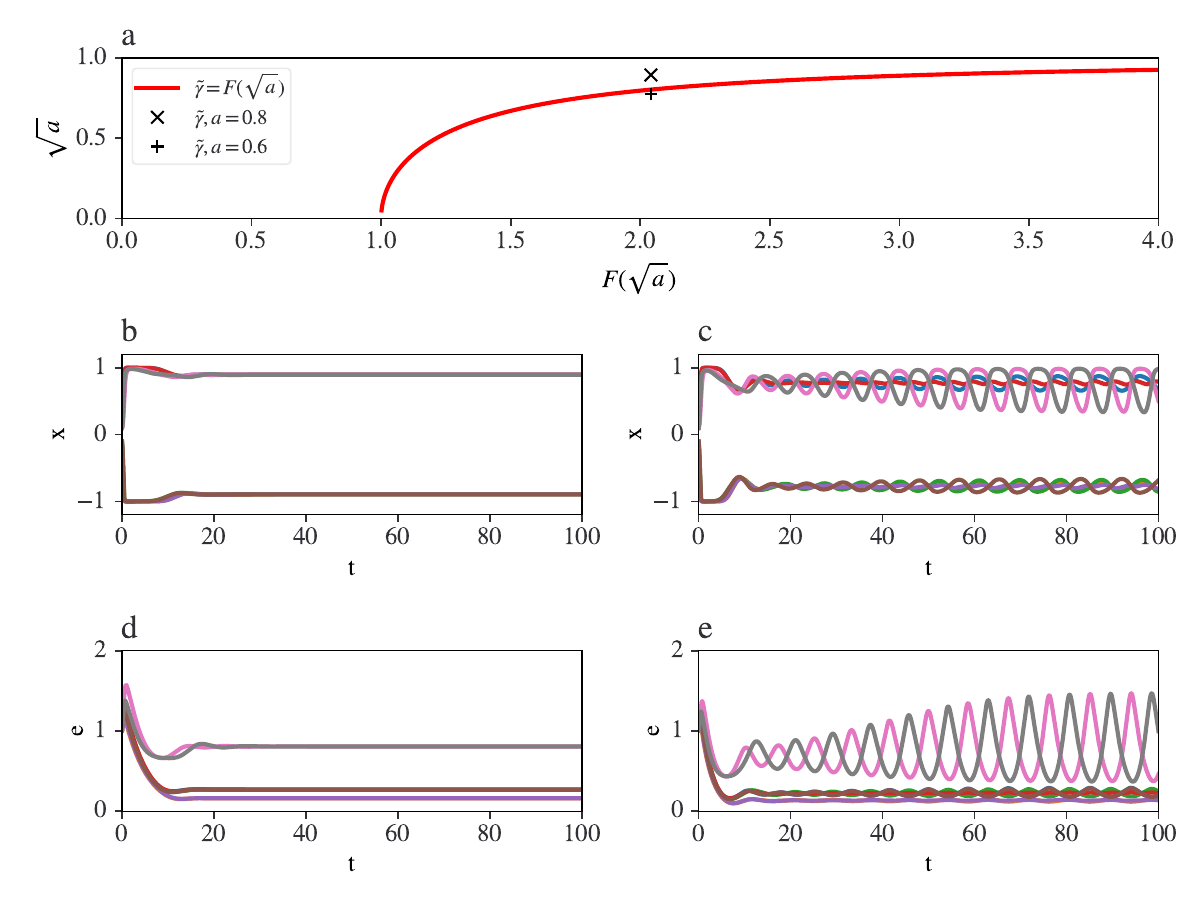}
  \caption{\label{fig:S2}(a) Bifurcation diagram of the system of eqs.~\ref{eq:S1} and \ref{eq:S2} in the case of the $V(\boldsymbol{x}) = -\frac{1}{2} \sum_{ij} \omega_{ij} \sigma_i \sigma_j$ with $N=8$ and $\omega_{ij}$ is randomly chosen to be equal to $1$ or $-1$ for $j>i$. $\omega_{ij} = \omega_{ji}$ and $\omega_{ii} = 0$. The red curve corresponds to the set $\{\sqrt{a},F(\sqrt{a})\}$ with $a>0$ which denote the condition under which the real part of the eigenvalues $\mu$ of the Jacobian matrix $J_{xx}$ is zero. The $+$ and $\times$ markers correspond to the maximum real part $\tilde{\gamma}$ of the eigenvalues of $D[\frac{\sigma_i}{\frac{\partial V}{\partial x_i}(\sigma_i \sqrt{a})}] H(\sqrt{a}\sigma_i)$ for the x-axis and the value of $a=0.6$ and $a=0.8$, respectively. (b) and (c) Time series of soft spins $x_i$ vs. time in the case $a=0.8$ and $a=0.6$, respectively. (d) and (e) The same as (b) and (c) for the error variables $e_i$.}
\end{figure}

In Fig. \ref{fig:S2}, the stability of the dynamics described in eqs. \ref{eq:S1} and \ref{eq:S2} when $V$ is Ising Hamiltonian with $N=8$ is shown by the bifurcation diagram in the space $\{F(\sqrt{a}),\sqrt{a}\}$. The $+$ and $\times$ markers correspond to the position real part of the eigenvalue of $\tilde{\gamma}$ of $D[\frac{\sigma_i}{\frac{\partial V}{\partial x_i}(\sigma_i \sqrt{a})}] H(\sqrt{a}\sigma_i)$ for the x-axis and the value of $a=0.6$ and $a=0.8$, respectively. For $a=0.8$, the point $\{\tilde{\gamma},\sqrt{a}\}$ is above the set defined by $\{\sqrt{a},F(\sqrt{a})\}$ with $a>0$ (shown in red). In this regime, the system settles to a stable fixed point as shown in Fig. \ref{fig:S2} (b) and (d). For $a=0.6$, the point $\{\tilde{\gamma},\sqrt{a}\}$ is below the red curve. In this regime, the system does not possess any fixed points and the limit behavior is a periodic cycle.

\newpage

\section{Pseudo-code of MHCACm}

The Pseudo-code of Metropolis-Hastings adjusted chaotic amplitude control with momentum (MHCACm) is shown in this section.

\begin{algorithm}[H]
\label{algo:pseudo}
\caption{Metropolis-Hastings adjusted chaotic amplitude control with momentum}
\begin{algorithmic}[1]
\footnotesize
\Function{Deterministic path}{$\sigma$} \Comment{Iterate deterministic n steps}
\State $u \gets  0$
\State $u' \gets  0$
\State $u'' \gets  0$
\State $e \gets 1$
\State $P \gets \frac{\sigma+1}{2}$ \Comment{Initialize}
\For{$t \gets 0$ to $n$}
\State $u'' \gets  u'$
\State $u' \gets  u$
\State $\mu \gets \nabla V_x(2P-1)$ \Comment{Calculate coupling}
\State $u \gets u' - \alpha u' + e \circ \mu + \gamma (u' - u'')$ \Comment{Euler steps}
\State $e \gets e - \xi ((2P-1) \circ (2P-1) - a) \circ e$
\State $P \gets \frac{1}{1+e^{- 2 \tilde{\beta} u}}$
\State $e \gets \frac{e}{<e>}$ \Comment{Normalize $e$}
\EndFor
\State \textbf{return} $x$ \Comment{Return the result}
\EndFunction

\Function{Probabilistic jump}{$x^2$} \Comment{Generate random jump}

\State $ \sigma^2 = 1$ with probability $P \gets \phi(2 \tilde{\beta} u/(\alpha e))$, $ \sigma^2 = -1$ otherwise \Comment{Random jump}

\State \textbf{return} $\sigma^2$ \Comment{Return the result}

\EndFunction

\Function{Metropolis-Hastings step}{$(\sigma^2,\sigma^1,x^2 ,\tilde{x}^1)$}  \Comment{Metropolis-Hastings step}

\State $A \gets A(\sigma^2/\sigma^1) = \text{min}(1,\frac{P(\sigma^2) Q(\sigma^1/ \sigma^2)}{P(\sigma^1) Q(\sigma^2 / \sigma^1)})$ \Comment{See eqs}

\State \textbf{return} $\sigma^2$ \Comment{Return the result}

\EndFunction

\Procedure{Sample }{$V$}

\State $ x^2 \gets \mathcal{U}(0,1)$ \Comment{Random initialization}

\For{$k \gets 0$ to $K$} \Comment{Generate $K$ samples}

\State $ \sigma^2 \gets $ \MakeUppercase{Probabilistic jump}  $(x^2)$ 

\State $ \tilde{x}^1 \gets $ \MakeUppercase{Deterministic path}  $(\sigma^2)$ 

\State $ x^2, H^2 \gets $ \MakeUppercase{Metropolis-Hastings step}  $(\sigma^2,\sigma^1,\tilde{x}^2 ,\tilde{x}^1)$ 

\State List of samples updated with $\sigma^2$

\EndFor

\EndProcedure
\normalsize
\end{algorithmic}
\end{algorithm}

\clearpage

\section{Boltzmann sampling}

\subsection{Verification of Boltzmann sampling at equilibrium}

\begin{figure}[H]
  \centering
\includegraphics[width=1.00\columnwidth]{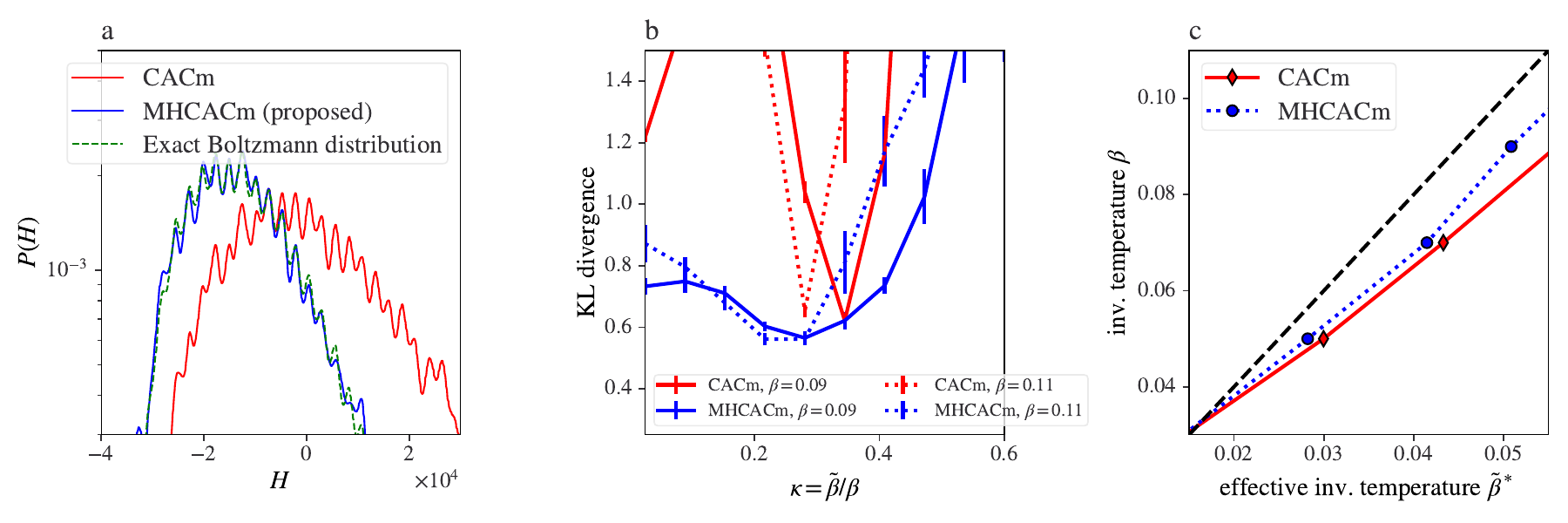}
  \caption{\label{fig:S2Boltz} (a) Exact Boltzmann distribution (green dotted line), sampled distribution using CACm (without MH step, see red line), and MHCACm (without MH step, see blue line). Gaussian filtering of the energy distribution is applied for the sake of readability. $\beta=0.07$. (b) KL divergence vs. ratio of inverse temperatures $\kappa = \frac{\tilde{\beta}}{\beta}$. (c) Effective inverse temperature $\tilde{\beta}^*$ temperature at which the KL divergence is minimized vs. $\beta$. The states used for the sampled distribution are boolean configurations accepted by the MH criterion.}
\end{figure}

To test our Markov chain approach, we sample from the Boltzmann distribution of the Ising Hamiltonian $V(\boldsymbol{\sigma}) = - \frac{1}{2} \sum_{ij} \omega_{ij} \sigma_i \sigma_j$. We use the Wishart planted ensemble (WPE)~\citep{hamze2020wishart} of size $N=18$ spins to generate the Ising coupling matrix. Using the WPE to generate instances of the Ising Hamiltonian allows us to tune the complexity of the energy landscape and also know the ground state a-priori~\citep{hamze2020wishart}. Additionally, these instances are known to exhibit properties of NP-Hard problems making the ground state difficult to find by heuristic algorithms~\citep{hamze2020wishart}. Because we use a small problem size ($N=18$) the exact Boltzmann distribution can be determined by brute force. The sampled distribution obtained by MHCACm is compared to the exact distribution at inverse temperature $\beta = 0.07$ by calculating the symmetric KL divergence (see Appendix section~\ref{sec:KLdef}). Note that CACm (without the Metropolis-Hastings step) does not sample from the target Boltzmann distribution (see Fig. \ref{fig:S2Boltz} a). In contrast, MHCACm exhibits exactly the same distribution of energy as the Boltzmann distribution at thermal equilibrium. This is accomplished while retaining the global exploration properties of CACm, making it a potent hybrid heuristic for identifying lower energy states of non-convex energy functions.

In order to evaluate the sampling characteristics of the constructed Markov chain, we aim at drawing samples from the Boltzmann distribution of the Ising Hamiltonian $V(\boldsymbol{\sigma}) = - \frac{1}{2} \sum_{ij} \omega_{ij} \sigma_i \sigma_j$ for a Wishart planted instance~\citep{hamze2020wishart} with a size of $N=18$ spins at an inverse temperature of $\beta$.

In Figs. \ref{fig:S2Boltz} (b) is shown that the KL divergence is minimized for a ratio of inverse temperatures $\kappa = \frac{\tilde{\beta}}{\beta}$. Note that $\tilde{\beta}$ is the effective temperature defined in eq. 2 of the main manuscript. The effective inverse temperature $\tilde{\beta}$ at which the KL divergence is minimized temperature in noted $\tilde{\beta}^*$. Figure \ref{fig:S2Boltz} (c) shows that $\tilde{\beta}^*$ is a linear function of $\beta$ within the range of parameters considered. Moreover, the addition of the MH step to MHCACm increases the similarity between $\beta^*$ and $\beta$ compared to that of CACm.

\subsection{Simulations for small $N$\label{sec:KLdef}}

In Fig. \ref{fig:S2Boltz}, the distance to the Boltzmann distribution $P(\boldsymbol{\sigma}) = \frac{e^{\beta V(\boldsymbol{\sigma})}}{Z}$ ($Z = \sum_{\boldsymbol{\sigma}} e^{\beta V(\boldsymbol{\sigma})}$) is calculated using the symmetric KL divergence. The symmetric KL divergence $D_{\text{SKL}}(P \parallel Q)$ between the distributions $P$ and $Q$ is defined as follows:

\begin{align}
D_{\text{SKL}}(P \parallel Q) = \frac{1}{2} (D_{\text{KL}}(P \parallel Q) + D_{\text{KL}}(Q \parallel P)), \label{eq:S26}
\end{align}

\noindent where

\begin{align}
D_{\text{KL}}(P \parallel Q) = \sum_{i} P(i) \log \frac{P(i)}{Q(i)}. \label{eq:S27}
\end{align}

In the case of Fig. \ref{fig:S2Boltz}, the convergence of the MCMC is measured by the KL divergence between the exact Boltzmann distribution of energy levels obtained by brute force calculations and the estimated distribution of energy levels obtained by simulation of the Metropolis-Hastings adjusted chaotic discrete space sampling. The former computation is possible because the problems size $N=18$ is sufficiently small for the brute force search method. The instance used for Fig. 2 is a randomly generated instance from the Wishart planted set with hardness parameter $\alpha_{\text{WPE}}=0.8$~\citep{perera2020chook}.

\subsection{Simulations for large $N$}

\begin{figure}[h]
  \centering
\includegraphics[width=1.00\columnwidth]{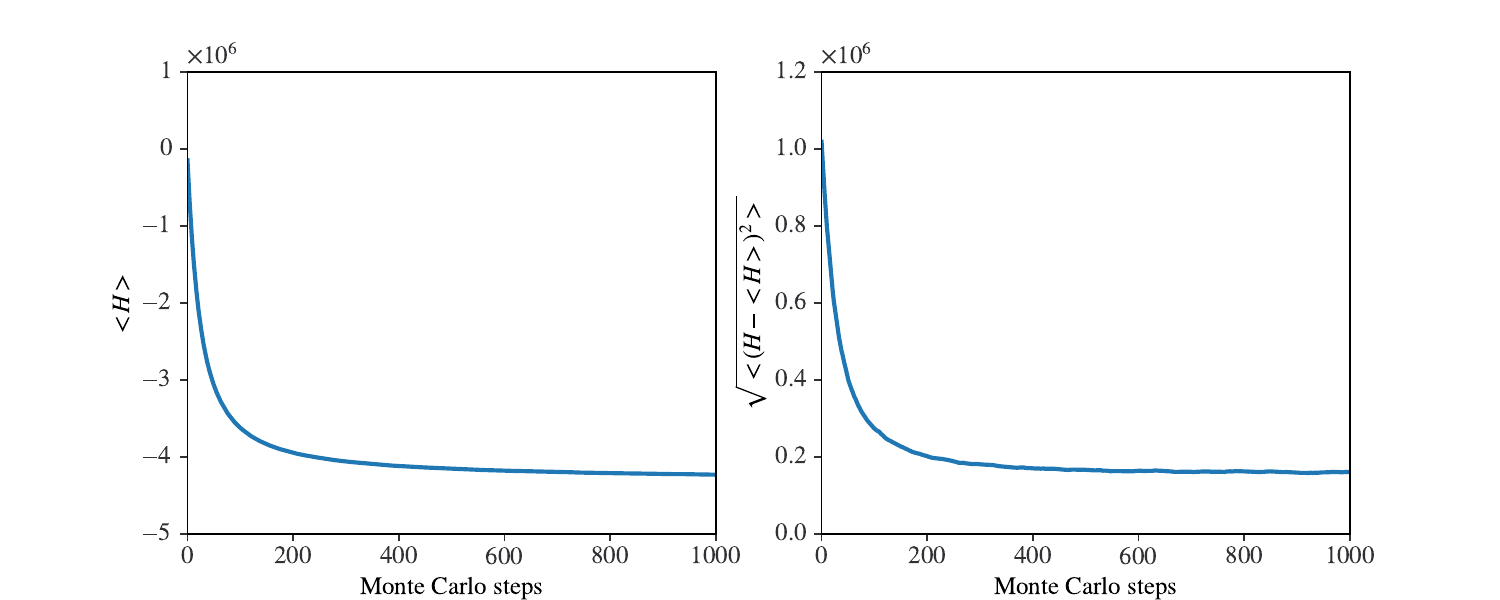}
  \caption{\label{fig:S3} Convergence of the mean and standard deviation of energy distribution of a $N=60$ Wishart planted instance with $\alpha_{\text{WPE}}=0.8$ under Gibbs sampling. Approximate convergence is reached after $1000$ Monte Carlo steps. $\beta=2.5$.}
\end{figure}

In order to test the convergence of the MCMC in the case of larger $N$, we can utilize a more heuristic metric. Conventional Gibbs sampling algorithm is run for a sufficiently large number of steps for the chain to converge. In the case of $N=60$ and $\alpha_{\text{WPE}}=0.8$, numerical simulations indicate that the convergence occurs within about 1000 time step of the Gibbs sampling algorithm. To observe the convergence, we limit the characterization of the Boltzmann distribution to its first two moments: the mean energy $<H>$ with $<H> = \sum_{\boldsymbol{\sigma}} H(\boldsymbol{\sigma}) P(\boldsymbol{\sigma})$ and standard deviation of the energy $\sqrt{<(H-<H>)^2>}$.  Figure \ref{fig:S3} shows the convergence of the mean and standard deviation of the energy under Gibbs sampling.

The distance D between the mean and variance of the distribution obtained by the proposed Metropolis-Hastings adjusted chaotic discrete space sampling and the one obtained by iterating the Gibbs sampling algorithm is then calculated as follows:

\begin{align}
D &= |<H_{\text{chaos}}>-<H_{\text{Gibbs}}>| \nonumber \\
&+ |\sqrt{<(H_{\text{chaos}}-<H_{\text{chaos}}>)^2>}-\sqrt{<(H_{\text{Gibbs}}-<H_{\text{Gibbs}}>)^2>}|,\label{eq:S28}
\end{align}

\noindent where the subscript denote the cases of Gibbs and MH chaotic sampling, respectively.

\begin{figure}[h]
  \centering
\includegraphics[width=1.00\columnwidth]{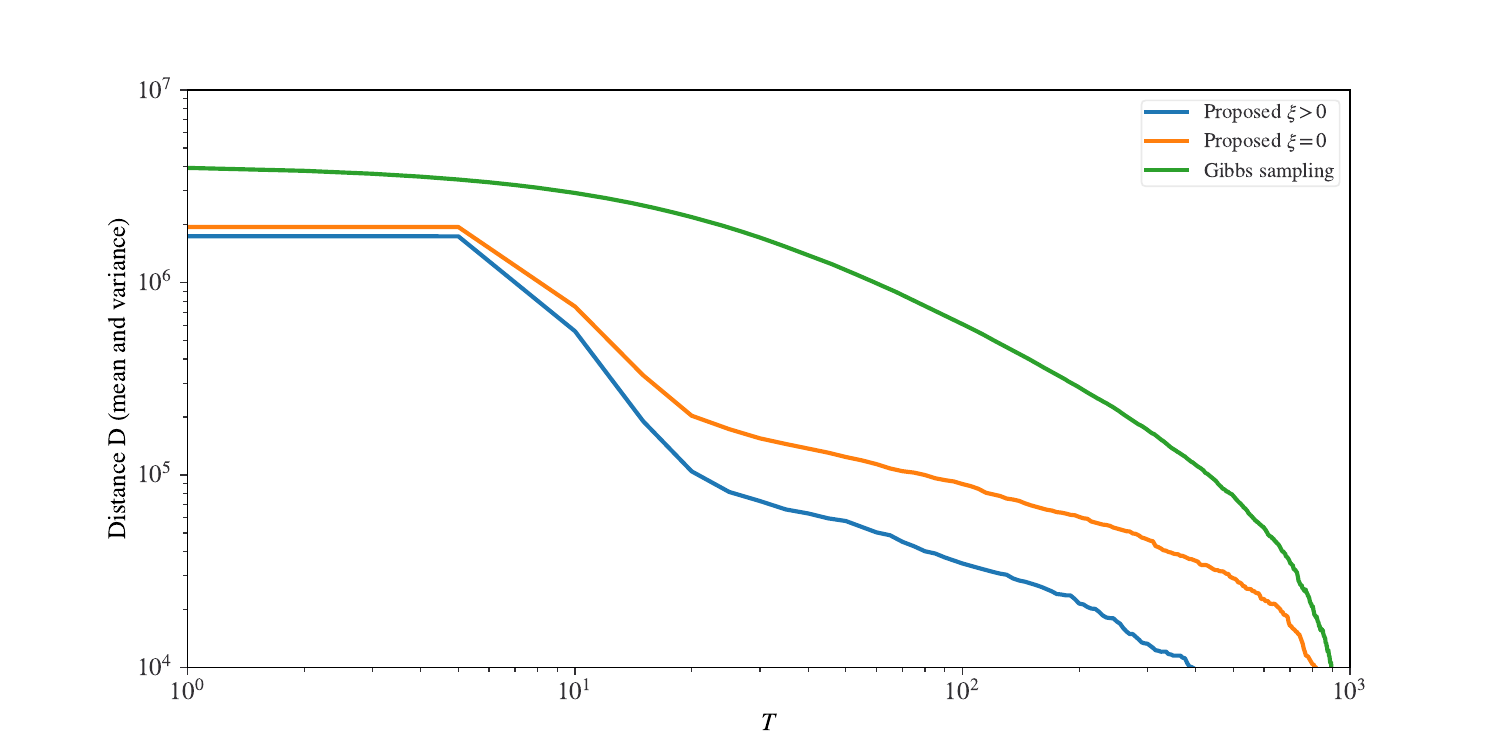}
  \caption{\label{fig:S4} Distance D between the mean and variance of the distribution obtained by the proposed Metropolis-Hastings adjusted chaotic discrete space sampling and the one obtained by iterating the Gibbs sampling algorithm (see eq.~\ref{eq:S26}) vs. the number of steps $T$ in the algorithm. One step corresponds to one tentative spin flip and one Euler step in the Gibbs sampling and chaotic sampling case, respectively.}
\end{figure}

In Fig. \ref{fig:S4} is shown distance $D$ vs. the number of steps $T$ in the algorithm. One step corresponds to one tentative spin flip and one Euler step in the Gibbs sampling and chaotic sampling case, respectively. The proposed sampling method exhibits a faster decrease of $D$ indicating a quicker relaxation to the steady-state distribution. The use of auxiliary variables speed up the convergence compared to the case without.

For $N$ larger than $N=60$, measuring the convergence of the distribution to its steady state becomes non-trivial because of strong broken ergodicity~\citep{Bernaschi2020}. Numerical simulation of the convergence to the steady state distribution in this case falls out of the scope of this paper. Note that the convergence to the steady-state distribution can be further accelerated by considering annealing of temperature or the use of the replica exchange Monte Carlo method in both cases of Gibbs sampling and proposed chaotic sampling.
\newpage

\section{Autotuning using dynamic anisotropic smoothing}

The parameters of the solvers are described in Table \ref{tab:params}. We have run the autotuning algorithm called dynamic anisotropic smoothing (DAS)~\citep{Reifenstein2024} for each problem setting (problem size $N$, bias $b$). The MH sampling rate is set to $f_{\text{MH}}=0.1$.

\begin{table}[H]
  \caption{Parameters of the algorithms tuned using DAS\label{tab:params}}
  \label{sample-table}
  \centering
  \begin{tabular}{l|l|l}
    \toprule
    Algorithm & number of parameters & parameter list \\
    \midrule
    SA & 3 & $\beta_{\text{ini}}$, $\beta_{\text{fin}}$, $T$\\
    AIM & 4 & $\beta$, $\alpha$, $\gamma$, $T$\\
    CAC & 5 & $\beta$, $\alpha$, $\xi$, $a$, $T$\\
    CACm & 6 & $\beta$, $\alpha$, $\xi$, $a$, $\gamma$, $T$\\
    HMCACm & 7 & $\beta$, $\kappa$, $\alpha$, $\xi$, $a$, $\gamma$, $T$\\
    \bottomrule
  \end{tabular}
\end{table}

\begin{figure}[h]
  \centering
\includegraphics[width=1.00\columnwidth]{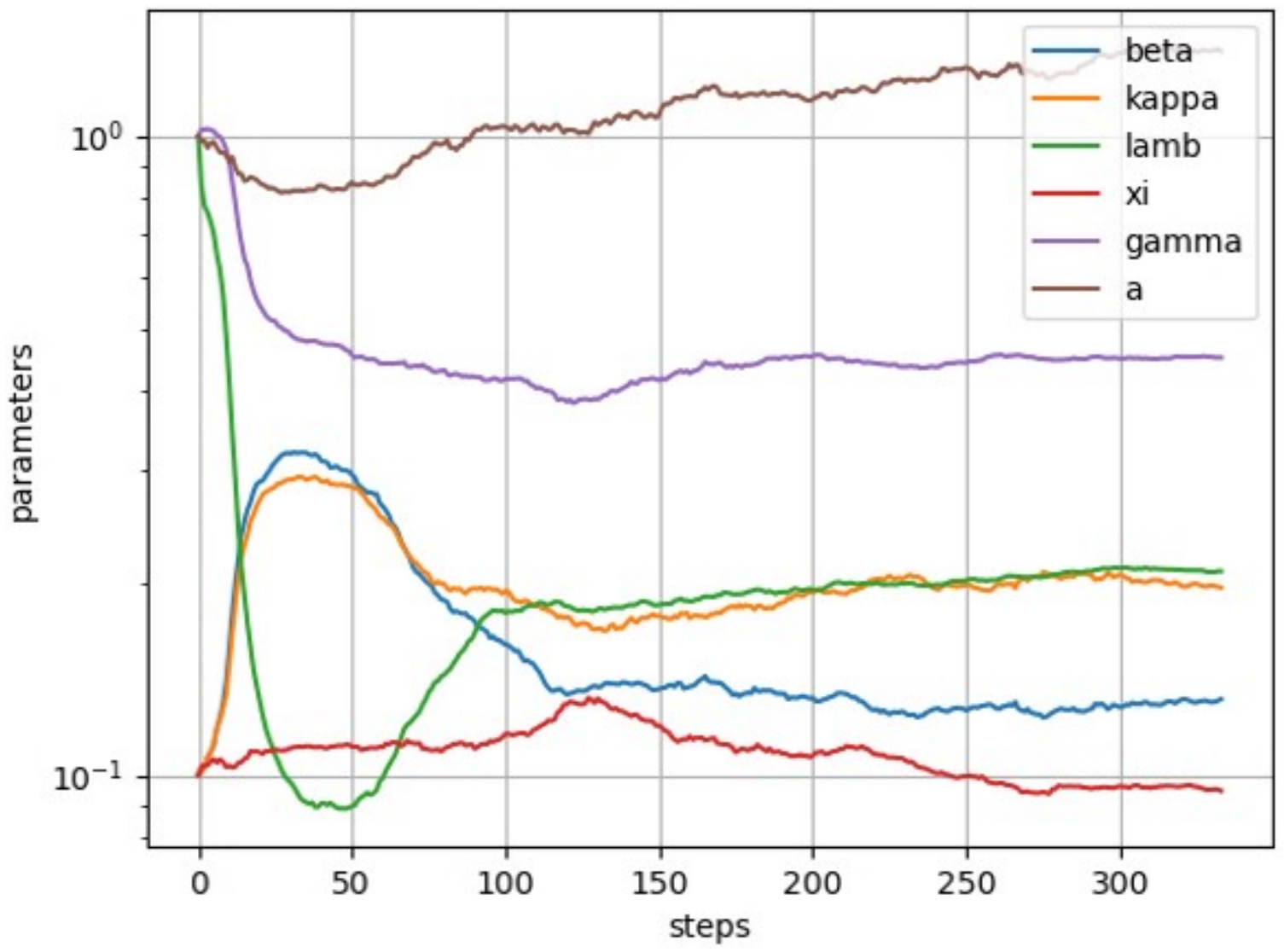}
  \caption{\label{fig:Sautotune} Evolution of parameters during the DAS autotuning for MHCACm. $N=100$, $b=12$, $T=1000$. Here ``lamb'' denotes $\alpha$ and ``kappa'' is $\frac{\tilde{\beta}}{\beta}$ }
\end{figure}

An example of the the evolution of parameters during the DAS autotuning is shown in Fig. \ref{fig:Sautotune}.

\newpage

\section{Contruction of degenerate Wishart planted ensemble}

The Wishart planted ensemble instances are generated by first choosing a planted ground state $g \in \{-1,1\}^N$. However, in our case we consider a set of planted ground states which are represented as columns in an $N$ by $D$ matrix $G \in \{-1,1\}^{N \times D}$. Let $\hat{G}$ be a matrix with the same columns space as $G$ whose columns are orthonormal. Then we generate an $N$ by $M$ matrix $\hat{W}$ of i.i.d. Gaussian random variables. This matrix is then transformed so that the columns are orthogonal to all of the ground states to form the matrix given as follows: $W = (I - \hat{G}\hat{G}^{\top})\hat{W}$. The Ising coupling matrix is constructed as $\hat{J} = WW^{\top}$ with the diagonal elements being set to zero. We can see that every column of $G$ is a ground state since $x^{\top} \hat{J}x \geq 0$ and $G^{\top}\hat{J}G = 0$. 
\\
\\
To add bias to the energy landscape, we want to somehow create the instance such that configurations close to one ground state tend to have a larger energy than the other ground state even though both ground states have the same energy. To do this, we can consider $W = (I - \hat{G}\hat{G}^{\top})A\hat{W}$ where $A$ is a symmetric matrix and $\hat{W}$ is the random matrix defined above. In this work we consider $A$ to be defined as $A = I + \frac{b}{N} \textbf{1}\textbf{1}^{\top}$ where $b$ is the "bias" away from the ferromagnetic solution. The point of this is that the residual energy (energy away from the ground state) will be amplified for states that are close to the ferromagnetic solution. Then, when choosing the planted ground state we can choose one which is close to the ferromagnetic solution and one that is far creating bias in the search towards one ground state over another. For the numerical results in this work we consider one ground state with a hamming distance of 3 from the ferromagnetic solution and one ground state chosen completely randomly from $ \{-1,1\}^N$. A Gauge transform is applied to obfuscate the ground states at the end of this process.

\section{Computing environment}

All simulations are done in CPU using a Intel Core i9-11950H, 8 cores, 2.6 GHz and GPU using a NVIDIA RTX A300; except for Table 3 of the main manuscript for which the GPU used is a Nvidia V100 and the CPU wall time is calculated on a 2024 MacBook Pro (Apple M3 Chip). The code is written in python and pytorch.

\end{document}